\documentclass[aps,prd,twocolumn,epsf,nofootinbib,superscriptaddress,showpacs]{revtex4}

\usepackage{graphicx,here}
\usepackage{amsmath,amssymb,latexsym}
\usepackage{bm}
\usepackage{url}

\def\lsim{\mathrel{\raise.3ex\hbox{$<$\kern-.75em\lower1ex\hbox{$\sim$}}}}
\def\gsim{\mathrel{\raise.3ex\hbox{$>$\kern-.75em\lower1ex\hbox{$\sim$}}}}

\def\lbldef#1#2{\expandafter\gdef\csname #1\endcsname {#2}}

\def\href#1#2{#2}

\newcommand{\bwide}{\begin{widetext}}
\newcommand{\ewide}{\end{widetext}}
\newcommand{\beq}[1]{\begin{equation} \label{(#1)}}
\newcommand{\eeq}{\end{equation}}
\newcommand{\ba}[1]{\begin{eqnarray} \label{(#1)}}
\newcommand{\ea}{\end{eqnarray}}

\begin{document}


\title{Are Lines From Unassociated Gamma-Ray Sources Evidence For Dark Matter Annihilation?}

\author{Dan Hooper}
\affiliation{Center for Particle Astrophysics, Fermi National Accelerator Laboratory, Batavia, IL 60510, USA}
\affiliation{Department of Astronomy and Astrophysics, The University of Chicago, Chicago, IL  60637, USA} 
\author{Tim Linden}
\affiliation{Department of Physics and Santa Cruz Institute for Particle Physics, University of California, 1156 High Street, Santa Cruz, CA 95064, USA}

\begin{abstract}
Very recently, it was pointed out that there exists a population of gamma-ray sources without associations at other wavelengths which exhibit spectral features consistent with mono-energetic lines at energies of approximately 111 and 129 GeV. Given recent evidence of similar gamma-ray lines from the Inner Galaxy, it is tempting to interpret these unassociated sources as nearby dark matter subhalos, powered by ongoing annihilations. In this paper, we study the spectrum, luminosity, and angular distribution of these sources, with the intention of testing the hypothesis that they are, in fact, dark matter subhalos. We find that of the 12 sources containing at least one prospective line photon, only 2 exhibit an overall gamma-ray spectrum which is consistent with that predicted from dark matter annihilations (2FGL J2351.6-7558 and 2FGL J0555.9-4348). After discounting the 10 clearly non-dark matter sources, the statistical significance of the remaining two prospective line photons is negligible. That being said, we cannot rule out the possibility that either or both of these sources are dark matter subhalos; their overall luminosity and galactic latitude distribution are not inconsistent with a dark matter origin. 
\end{abstract}

\pacs{95.35.+d;95.85.Pw;07.85.-m;98.70.Rz; FERMILAB-PUB-12-422-A}

\maketitle

\section{Introduction}

In many models, dark matter particles can annihilate through loop-level diagrams to final states which include a photon, such as $\gamma \gamma$, $\gamma Z$ or $\gamma h$. As dark matter annihilations in the universe today occur at non-relativistic velocities, the photons produced through such processes are predicted to take the form of mono-energetic gamma-ray lines. The prospect of observing such a gamma-ray line has long been considered a ``smoking gun'' for dark matter's indirect detection, providing a distinctive signal which is unlikely to be mimicked by astrophysical backgrounds.

The Fermi Gamma-Ray Space Telescope (Fermi-LAT)~\citep{atwood_fermi} is one of the most promising instruments with which to detect such a gamma-ray line. Over the past few months, a great deal of attention has been given to the possibility that evidence for such gamma-ray lines is, in fact, present within the publicly available data from the Fermi-LAT~\cite{Bringmann:2012vr,Weniger:2012tx}. The primary line in question appears at an energy of approximately 129 GeV, along with less statistically significant hints of a second line at around 111 GeV~\cite{Su:2012ft}. These photons have been detected primarily within regions of the sky which reside within $\sim$10$^{\circ}$ of the Galactic Center, resembling the distribution predicted from a cusped halo profile. For recent discussions of these possible signals and their implications for particle physics, see Refs.~\cite{Su:2012ft,data,dougidm,Boyarsky:2012ca} and~\cite{pp,Buckley:2012ws}, respectively.

The task now at hand for the particle-astrophysics community is to determine whether the gamma-ray line or lines observed from the Inner Galaxy are 1) actually the products of dark matter annihilations, 2) somehow the result of a combination of astrophysical sources or mechanisms~\cite{bubble,Aharonian:2012cs}, or 3) the result of systematic or instrumental issues~\cite{Su:2012ft,dougidm,Boyarsky:2012ca}, perhaps associated with the inner workings of the Fermi-LAT itself. At this time, it is difficult to take very seriously the possibility that the appearance of the 129 GeV line is a statistical fluctuation, in particular in light of the very high significance of the line as derived from the template analysis of Ref.~\cite{Su:2012ft} (greater than 5$\sigma$, after accounting for an appropriate trials factor).

One way to possibly strengthen the case for a dark matter origin of the observed line would be to observe a line at the same energy from other directions of the sky thought to contain significant densities of dark matter. Such a signal has been recently reported, for example, from a number of Galaxy Clusters, although with only modest statistical significance~\cite{Hektor:2012kc}. Even more recently, evidence of gamma-rays lines at both 129 and 111 GeV from a collection of unassociated gamma-ray sources has been presented~\cite{new}. If confirmed, this result could have great bearing on the question of the line's origin. The Milky Way is predicted to contain large numbers of smaller subhalos and nearby subhalos could plausibly appear as a population of gamma-ray sources without counterparts at other wavelengths~\cite{Belikov:2011pu,Mirabal:2012em,Buckley:2010vg,Zechlin:2011kk,acts,Kuhlen:2008aw,Zechlin:2011wa,Mirabal:2010ny}. If the line emission from such sources is confirmed, this would be most easily interpreted as evidence of dark matter annihilations taking place within such objects. 

In this paper, we study the gamma-ray spectrum, luminosity, and sky distribution of these prospective line-emitting sources with the goal of assessing the likelihood that they are in fact dark matter subhalos. If a sizable fraction of these sources were shown to be subhalo-like in these regards, it would help to strengthen the case that the gamma-ray lines reported over the past several months do, in fact, originate from dark matter annihilations. What we actually find, however, is that the majority of these objects do not appear to be dark matter subhalos. That fact that line emission is observed from these objects, which do not appear to be powered by dark matter annihilations, adds credibility to the hypothesis that the lines reported from the Inner Galaxy are the product of some yet unknown systematic or instrumental issue, possibly associated with the Fermi-LAT (see also Ref.~\cite{Su:2012ft,dougidm}).

\section{The Spectra of Line Emitting Candidate Sources}

In their recent study, the authors of Ref.~\cite{new} studied the subset of the unassociated sources within the Second Fermi Source Catlog (2FGL)~\cite{2catalog} which do not exhibit discernible variability and which lie outside of the Galactic Plane ($|b| > 5^{\circ}$). From among the hundreds of sources which meet these requirements, they found that 16 of them had at least one 100-140 GeV photon (within 0.15$^{\circ}$ or 0.3$^{\circ}$ of the source's center for front- or back-converting events, respectively). The stacked spectrum of these sources reveals peaks at around 129 and 111 GeV, consistent with the energies of gamma-rays tentatively observed from the Inner Milky Way. More specifically, they find 6 sources in the 2FGL catalog showing emission at energies associated with each of these two lines. The sources J0341.8+3148c, J2115.4+1213, J1716.6-0526c, J2351.6-7558, J1639.7-5504, and J2004.6+7004 are found to emit a photon in the energy range 124.7-133.4 GeV and the 2FGL sources J0555.9-4348, J1844.3+1548, J1240.6-7151, J1324.4-5411, J0600.0+3839, and J1601.1-4220 are found to emit at least one in the energy range range 108.9-116.6 GeV~\citep{finkbeiner_communication}. The statistical significance of the combined 129 and 111 GeV line-like emission (relative to the prediction of a conservative background model) from the unassociated sources was evaluated to be 3.3$\sigma$~\cite{new}. 

\begin{figure}[t]
\begin{center}
{\includegraphics[angle=0,width=0.9\linewidth]{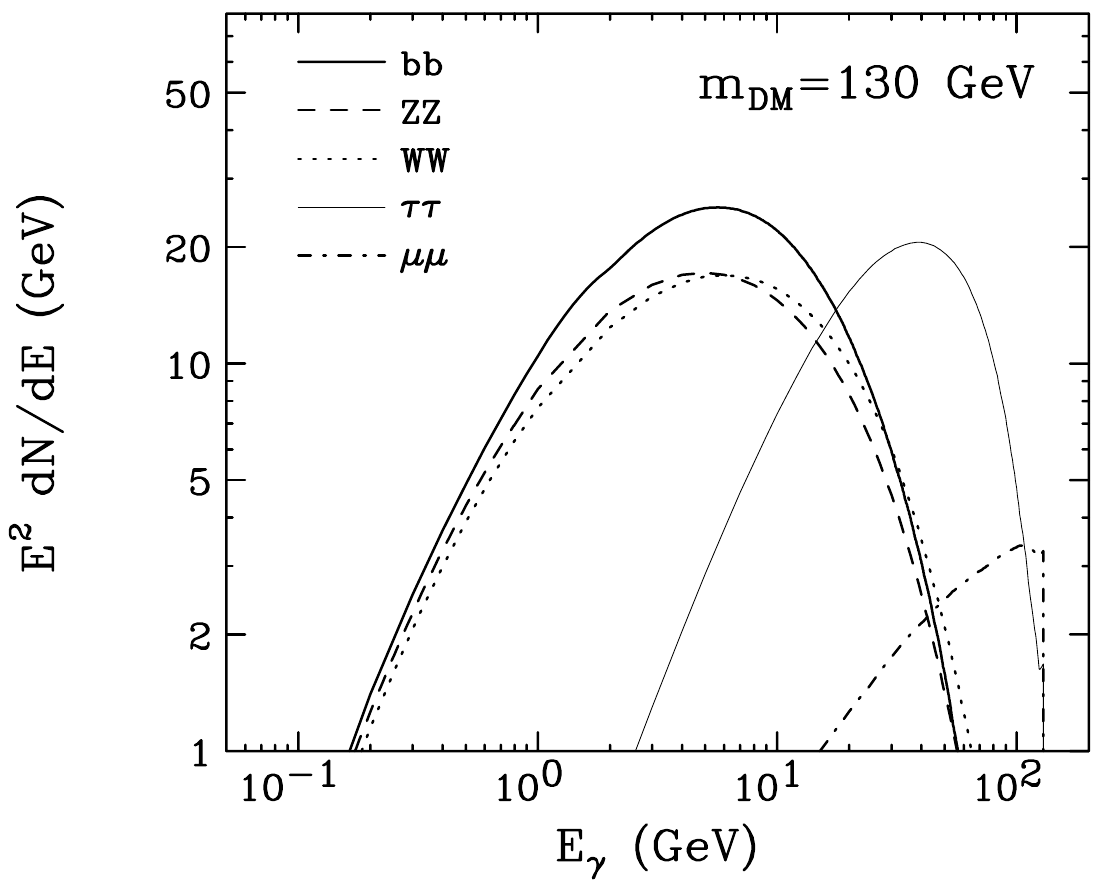}}
\vspace{-0.3cm}
\caption{The spectrum of gamma-rays per dark matter annihilation, for a number of channels and for a WIMP mass of 130 GeV.}
\label{spectra}
\end{center}
\end{figure}

\begin{figure*}[t]
\begin{center}
{\includegraphics[angle=0,width=0.87\linewidth]{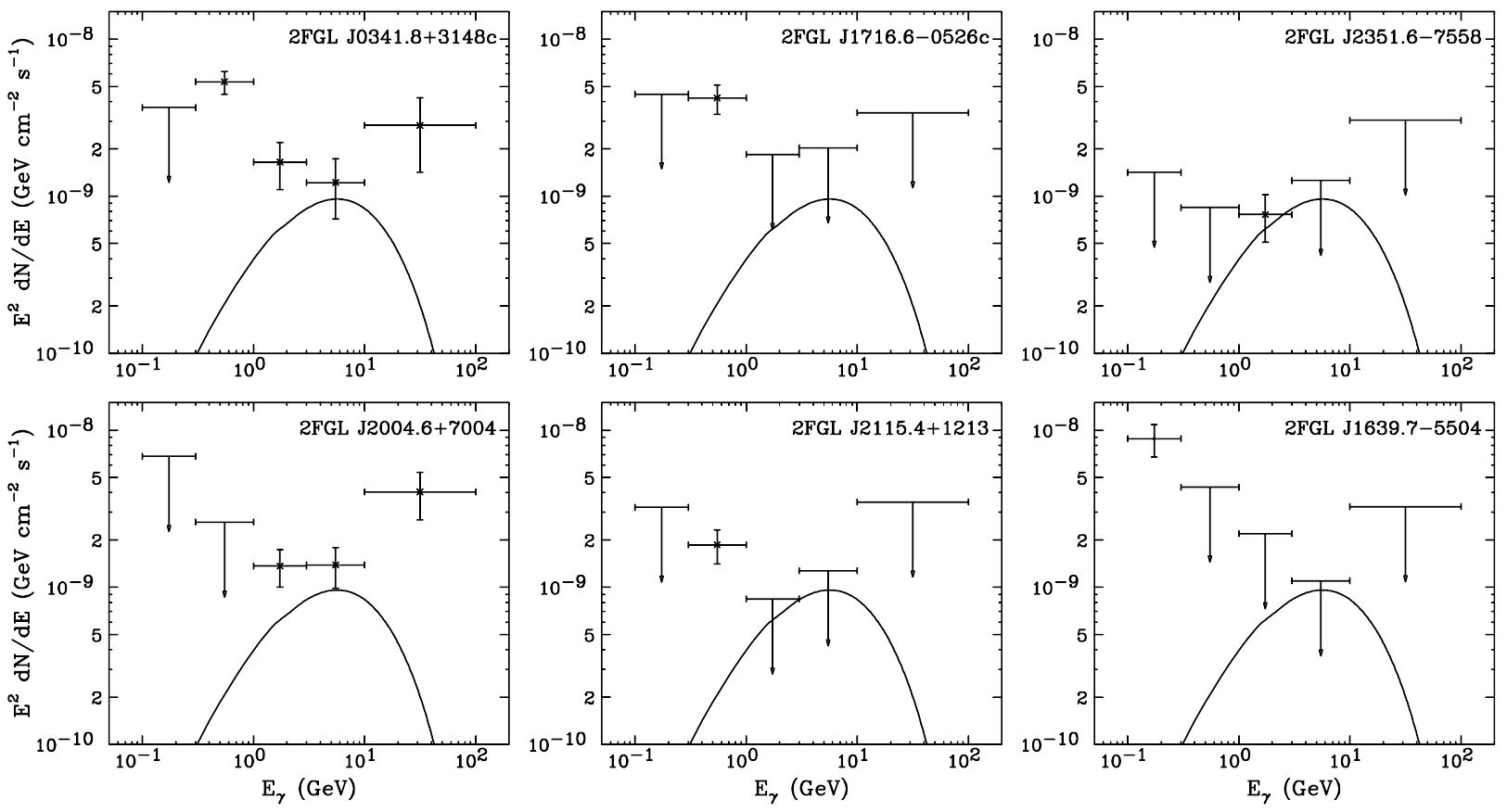}}
\vspace{-0.3cm}
\caption{The gamma-ray spectra from the six non-variable (variability index~$<$~41.6) and non-planar ($|b|>5^{\circ}$), unassociated sources as calculated in the Second Fermi Source Catalog (2FGL) which have had a photon detected with an energy between 124.7 and 133.4 GeV. Arrows represent 95\% upper limits, while other points denote 1$\sigma$ errors. In each frame, the solid line is the spectral shape (arbitrary normalization) predicted from 130 GeV dark matter annihilations to $b\bar{b}$. From among these six sources, only the spectrum of 2FGL J2351.6-7558 is consistent with that predicted from annihilations taking place in dark matter subhalos. We note that this conclusion is unchanged if annihilations through other channels are considered.\\}
\label{can1}
\end{center}
\end{figure*}
\begin{figure*}[!]
\begin{center}
{\includegraphics[angle=0,width=0.87\linewidth]{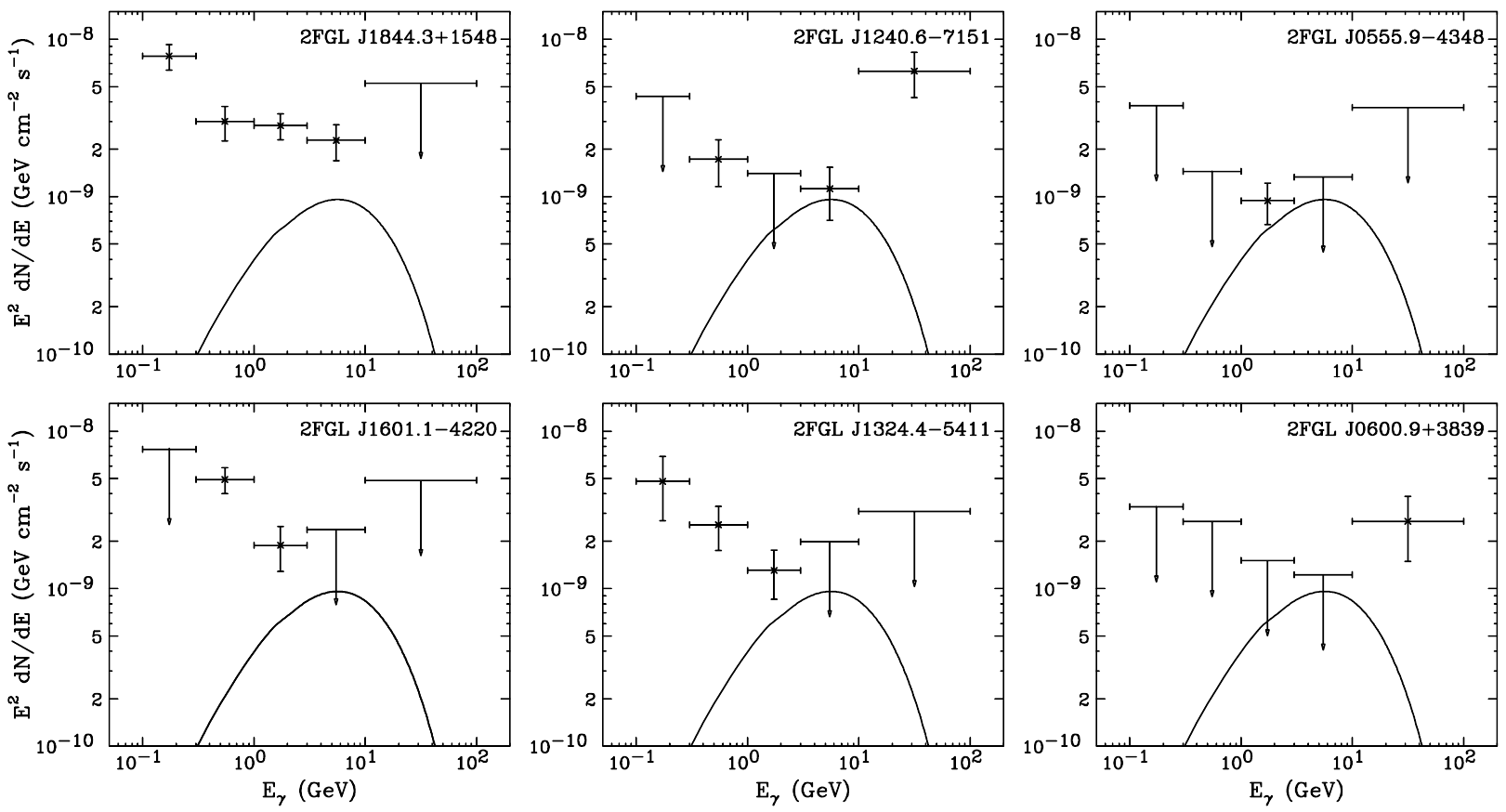}}
\vspace{-0.3cm}
\caption{The gamma-ray spectra from the six non-variable (variability index~$<$~41.6) and non-planar ($|b|>5^{\circ}$), unassociated sources as calculated in the Second Fermi Source Catalog (2FGL) which have had a photon detected with an energy between 108.9 and 116.6 GeV. Arrows represent 95\% upper limits, while other points denote 1$\sigma$ errors. In each frame, the solid line is the spectral shape (arbitrary normalization) predicted from 130 GeV dark matter annihilations to $b\bar{b}$. From among these six sources, only the spectrum of 2FGL J0555.9-4348 is consistent with that predicted from annihilations taking place in dark matter subhalos. We note that this conclusion is unchanged if annihilations through other channels are considered.}
\label{can2}
\end{center}
\end{figure*}

If any of these sources are dark matter subhalos, then in addition to any line emission, a continuum of gamma-rays is also expected to result from dark matter annihilations to final states other than $\gamma\gamma$. Furthermore, it is precisely this continuum emission that would lead to such a source appearing in the 2FGL (the line emission from such a source alone is not bright enough for Fermi to detect at high significance). In Fig.~\ref{spectra}, we plot the spectrum of continuum gamma-rays from a 130 GeV dark matter particle annihilating to several possible final states (as calculated using PYTHIA 6~\cite{pythia}). Most of the final states common to the most studied and well motivated WIMP candidates (quarks, gauge bosons) yield a spectrum of gamma-rays which peaks at around $\sim$5-6 GeV (in $E^2 dN/dE$ units). Annihilations to tau-lepton pairs, in contrast, yield a larger fraction of hard neutral pions, and thus result in a gamma-ray spectrum which peaks at higher energies, around $\sim$40 GeV. Annihilations to light leptons (muons or electrons) produce gamma-rays primarily through final state radiation, resulting in a spectrum which peaks at even higher energies.

By comparing the spectra of the unassociated sources discussed in Ref.~\cite{new} to that predicted from dark matter continuum emission, it should be possible to test the hypothesis that the observed line emission is associated with dark matter annihilations. In Figs.~\ref{can1} and \ref{can2}, we show the spectrum of each of the 12 sources containing a photon consistent with the gamma-ray lines, as contained in the 2FGL catalog, and compare this spectral shape with that predicted from a 130 GeV dark matter particle annihilating to $b\bar{b}$ (the normalization of the annihilation spectrum is arbitrary in these two figures; an issue we will return to in the next section). For 10 of these 12 sources, the spectral shape is clearly incompatible with the continuum emission predicted from dark matter annihilation. In particular, only 2FGL J2351.6-7558 and 2FGL J0555.9-4348 exhibit spectra which could plausibly originate from the annihilations of a 130 GeV dark matter particle. Other annihilation channels, such as those to gauge bosons or leptons, do not fit this collection of sources any better. We also note that the spectra of the two sources which are potentially compatible with a dark matter interpretation are not particularly well measured, having fluxes inconsistent with zero only in the 1-3 GeV band. In light of this, we find little to support the conclusion that dark matter annihilations are responsible for the gamma-ray lines identified in Ref.~\cite{new}.

\clearpage

\begin{figure*}[t]
\begin{center}
{\includegraphics[angle=0,width=0.7\linewidth]{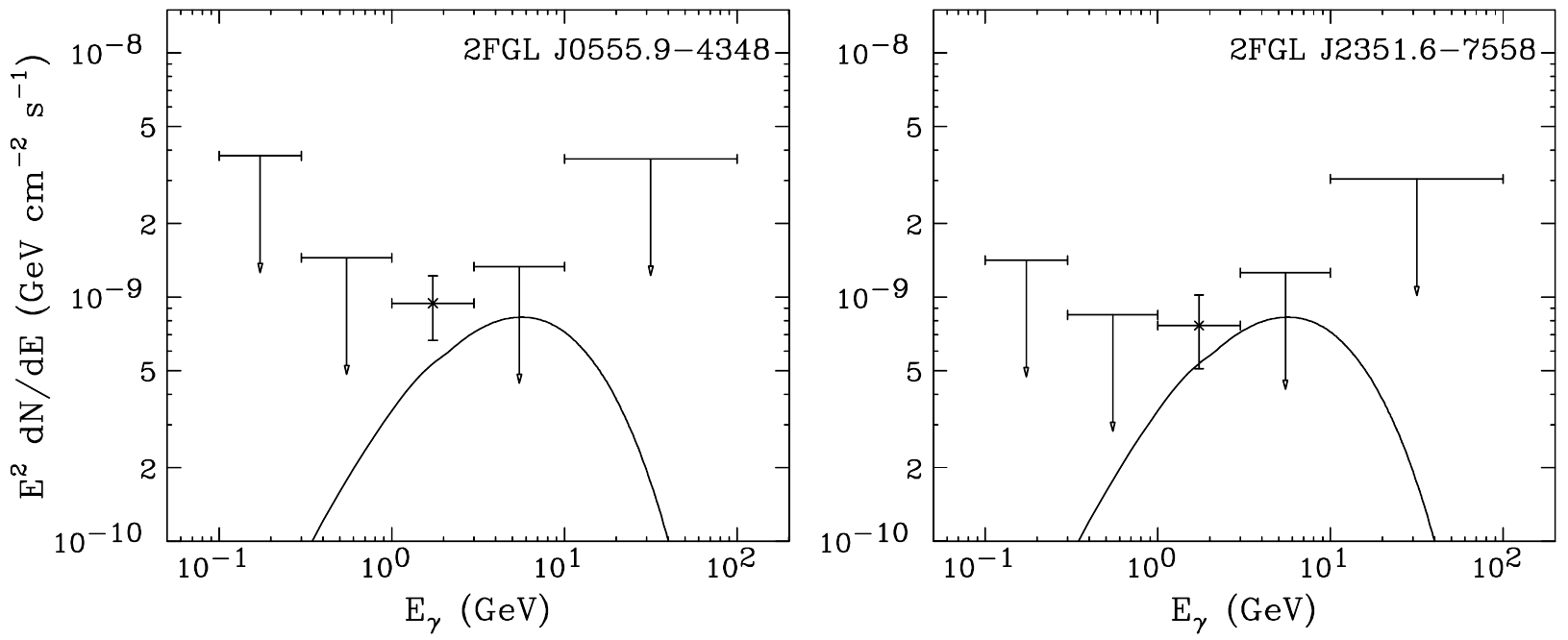}}
\vspace{-0.3cm}
\caption{The maximum spectrum of gamma-ray continuum emission from the two line-emitting source candidates in the 2FGL with spectral shapes consistent with originating from dark matter (for the case of annihilations to $b\bar{b}$, although results for annihilations to other quarks or gauge bosons are similar). If the flux of continuum emission was higher, it would exceed that observed from the Inner Galaxy~\cite{Buckley:2012ws}. This result assumes that the single line photon from each of these sources does not represent a significant upward or downward fluctuation.}
\label{norm}
\end{center}
\end{figure*}

\section{Is The Continuum Emission Too Bright?}

The luminosity of gamma-rays lines produced from dark matter annihilations taking place in a subhalo is given by:
\begin{equation}
L_{\rm line}=\frac{[2\sigma_{\gamma \gamma} v+\sigma_{\gamma Z}+\sigma_{\gamma h}]}{2 m^2_{\rm DM}} \int \rho^2 dV,
\end{equation}
where $\sigma v$ and $m_{\rm DM}$ are the annihilation cross sections and mass of the dark matter particle, respectively, and the integral is performed over the volume of the subhalo. 

In contrast, the luminosity of continuum gamma-rays is given by:
\begin{equation}
L_{\rm cont}=\frac{\sigma_{\rm cont} v}{2 m^2_{\rm DM}} \int \frac{dN_{\gamma}}{dE_{\gamma}} dE_{\gamma} \int \rho^2 dV,
\end{equation}
where $dN_{\gamma}/dE_{\gamma}$ is the spectrum of continuum gamma-rays produced per dark matter annihilation, which depends on the dominant annihilation channel(s) and on the mass of the dark matter particle. 

Combining these expressions, we can write the ratio of continuum-to-line emission from a given subhalo independently of the dark matter distribution:
\begin{equation}
\frac{L_{\rm cont}}{L_{\rm line}} = \frac{\sigma_{\rm cont} v}{[2\sigma_{\gamma \gamma} v+\sigma_{\gamma Z}v+\sigma_{\gamma h}v]} \, \int \frac{dN_{\gamma}}{dE_{\gamma}} dE_{\gamma}.
\end{equation}

Each of the 12 sources under consideration have only one or two photons with an energy consistent with annihilation to a 129 or 111 GeV gamma-ray line (all but one of these 12 sources has only one prospective line photon). If we set the quantities that determine $L_{\rm line}$ such that we expect one line photon from a given source, we can then use that information to predict the flux of continuum emission from that source, as a function of $L_{\rm cont}/L_{\rm line}$. An upper limit on the ratio of the line-to-continuum emission can be placed from the fact that little or no continuum emission from a 130 GeV WIMP has been observed from the regions of the Inner Galaxy where the line (or lines) have been observed. Comparing the cross section to gamma-ray lines found in Ref.~\cite{Weniger:2012tx} to the upper limits on continuum emission found in Refs.~\cite{Buckley:2012ws,Hooper:2011ti} (and adopting the same halo profile), we find that this requires $\sigma_{b\bar{b}} v/[\sigma_{\gamma \gamma} v+0.5\,\sigma_{\gamma Z}v+0.5\,\sigma_{\gamma h} v] < 6.9$. In Fig.~\ref{norm}, we compare this maximal prediction for the spectrum of continuum emission to that observed from 2FGL J2351.6-7558 and 2FGL J0555.9-4348. In each case, the maximum continuum spectrum predicted is lower than the observed flux, but by only about a factor of two. If the individual line photons observed from each of these two sources were downward fluctuations (on an expectation of two or more line photons from each source), then the required normalization could be consistent with constraints from the Inner Galaxy. If these two sources are in fact dark matter subhalos, they must be among the very brightest of all subhalos, in terms of both line and continuum emission.



\section{Is The Number of Line Emitting Sources Consistent With Subhalo Predictions?}

In this section, we calculate the number of dark matter subhalos predicted to be bright enough to be detectable as 2FGL sources, and compare this to the number of sources observed to possibly exhibit line emission.

Numerical simulations of the formation and evolution of structure have found that dark matter halos and subhalos follow a mass distribution approximately given by $dN_n/dM_h \propto M_h^{-2}$, extending down to a minimum mass related to the microscopic properties of the dark matter particle~\cite{smallest}. In the case of the Milky Way, on the order of 5-10\% (50\%) of the total mass in dark matter is predicted to be found in $10^7$-$10^{10} \, M_{\odot}$ ($10^{-6}$-$10^{10} \, M_{\odot}$) subhalos~\cite{norm}.  If the distribution of subhalos extends down to the Earth mass-scale, this normalization corresponds to a total of more than $10^{16}$ subhalos within the halo of the Milky Way, corresponding to a number density of approximately $\sim$$10^2$ subhalos per cubic parsec in the local neighborhood of the Galaxy. 

The gamma-ray luminosity from dark matter annihilations in a given subhalo depends on how the dark matter is distributed. We begin by considering an NFW halo profile:
\begin{equation}
\rho(r) \propto \frac{1}{(r/R_s)\, [1+(r/R_s)]^2},
\label{1pt2}
\end{equation}
where $R_s$ is the scale radius of the subhalo, which we set for a halo of a given mass according to the the analytic model of Bullock {\it et al.}~\cite{bullock}. Considerable halo-to-halo variation in the concentration and shape of subhalo profiles has been observed in numerical simulations. In contrast, the model of Ref.~\cite{bullock} only provides a measure of the average concentration of a subhalo of a given mass. We model the probability of a halo having a given concentration with a log-normal distribution with a dispersion of $\sigma_c\approx 0.24$~\cite{bullock}. Such variations roughly double the number of observable subhalos~\cite{Pieri:2007ir}. 

Subhalos in the local volume of the Milky Way are likely to have had a large fraction of their outer mass stripped through tidal interactions with other halos and stars, leaving a dense and tightly bound inner cusp intact~\cite{disruption}. As the innermost regions of halos dominate the overall annihilation rate, however, the precise fraction of mass that is lost only modestly impacts the resulting gamma-ray luminosity~\cite{loss}.

In order for the gamma-ray annihilation products from a given subhalo to constitute a source that could potentially appear within the 2FGL, the subhalo must be fairly bright. A subhalo that produces more than 50 events above 1 GeV per year at FGST should be likely to appear within the 2FGL~\cite{Belikov:2011pu}. 


In calculating our initial estimate for the number of dark matter subhalos bright enough to appear in the 2FGL, we assume that the outer 99\% of the original halo's mass is lost to tidal stripping, and integrate down to substructure masses of 10$^{-6}\, M_{\odot}$ (corresponding to a boost factor of 1.75). In this case, with a dark matter mass of 130 GeV, annihilating with an annihilation cross section of $\sigma v=8\times 10^{-27}$ cm$^3$/s (which saturates the Galactic Center constraints for the halo profile implied by the line morphology~\cite{Buckley:2012ws}), we find that there is only a 1-in-50 chance of Fermi observing even a single dark matter subhalo. 

If we consider a more optimistic, but still plausible, set of assumptions, this estimate can increase significantly. For example, if we repeat this calculating using a somewhat steeper dark matter distribution ($\rho \propto r^{-1.2}$ in the inner volume~\cite{1pt2}), only 95\% mass loss, and integrate substructure down to $10^{-8} \, M_{\odot}$ (corresponding to a boost factor of 2.8), we find that Fermi should have about a 70\% chance of observing at least one dark matter subhalo. In light of this result, we find it plausible that as many as one or two of the unassociated 2FGL sources could be subhalos of 130 GeV dark matter particles responsible for the Galactic Center line emission. We find the observation of 12 dark matter subhalos, however, relatively difficult to explain within the context of existing models of dark matter structure.


We note that if the dark matter profiles and substructure boost factors of dwarf galaxies were as favorable as the aforementioned optimistic values needed to yield observable subhalos, then we should also have expected to observe gamma-rays from these objects as well. Searches for continuum emission~\cite{fermi_dwarfs} and line emission~\citep{GeringerSameth:2012sr} from dwarf galaxies, however, have revealed no such signals. A natural resolution to this apparent conflict could be found if the dark matter distributions in dwarf galaxies are only mildly cusped, or even cored, perhaps as a result of baryonic effects~\cite{Brooks:2012vi} which are negligible in smaller subhalos.


\section{Is the Latitude Distribution of Line Sources Consistent with Expectations for Subhalos?}

As pointed out in Ref.~\cite{new}, the distribution of potentially line-emitting 2FGL sources is noticeably concentrated around the Galactic Plane. In particular, of the 12 sources under consideration, 6 (10) of them reside within the region of $5^{\circ} < |b| < 10^{\circ}$ ($5^{\circ} < |b| < 25^{\circ}$). Sources within $5^{\circ}$ of the Galactic Plane were not considered in Ref.~\cite{new}. This distribution of sources is strongly concentrated around the plane and is at odds with that expected for a population of dark matter subhalos.

Modern simulations of cold dark matter lead us to expect a large fraction of observable subhalos to reside significantly outside of the Galactic Plane~\citep{aquarius,vialactea}. We expect the nearby subhalos most likely to be observable by Fermi to follow a roughly isotropic distribution across the sky. Although arguments have been presented for why this isotropy may be broken, these lines of reasoning lead to the prediction that an even larger fraction of subhalos should reside well outside of the Galactic Plane.\footnote{The overall distribution of subhalos in the Milky Way is expected to be skewed by a preferred alignment of the subhalo population with the major axis of the triaxial mass distribution of their dark matter hosts~\citep{subhalo_anisotropy,subhalo_distribution_anisotropy}. This can be understood in the context of the accretion of satellite galaxies along a small number of dark matter filaments~\citep{accretion_from_filaments}. This axis is observed to be roughly perpendicular to the Galactic Plane, creating a relatively high latitude population of subhalos as observed from the solar position. This bias has been confirmed for the population of the 14 observed dark matter dwarfs, all of which have $|b|$~$>$~34$^\circ$, and 8 of which lie at $|b|$~$>$~50$^\circ$. On the other hand, recent simulations~\citep{simulations_anisotropy} reveal that this great pancake is likely to become less significant when increasingly less massive halos are included in the sample, likely due to the inclusion of subhalos from multiple accretion events. } In any case, one does not expect the distribution of dark matter subhalos to be concentrated at low galactic latitudes, as is observed among the 12 sources under consideration.

While these cold dark matter simulations do not include the highly non-spherical distribution of galactic baryons, the addition of the baryonic disk is thought to deplete dark matter substructure near the Galactic Plane, further increasing the number of high latitude halos with respect to those expected by dark matter only simulations~\citep{baryonic_depletion, ram_pressure}. Simulations by Ref.~\citep{anisotropy_of_dark_matter_galactic_plane} find that tidal disruptions of subhalos in the Galactic Plane allow for a 9\% anisotropy biasing the indirect detection of dark matter towards high latitudes. While the authors of Ref.~\citep{new} argue that the preferential observation of dark matter substructure at low galactic latitudes may be due to a selection effect, models employing dark matter substructure simulations predict the most luminous nearby halos to spread across a wide range of galactic latitudes~\citep{kuhlen_glast_predictions}.

If instead of considering all 12 potentially line-emitting sources, we limit ourselves to the two sources which are spectrally consistent with originating from dark matter annihilations, we come to a rather different conclusion. In particular, of the 12 sources, these two are the farthest away from the Galactic Plane, residing at galactic latitudes of -40.6$^\circ$ and -27.8$^\circ$. The latitude distribution of these two sources, although clearly statically insignificant, is not at odds with the predicted distribution of galactic subhalos.

\section{A Caveat? Inverse Compton Emission}


In principle, many of the arguments made in this paper concerning the spectrum and intensity of continuum emission can be evaded if these gamma-rays originate from secondary interactions, such as through bremsstrahlung with gas, or inverse Compton scattering with starlight. In order for these processes to dominate the observed gamma-ray flux, annihilations must proceed largely to gamma-ray quiet states, such as e$^+$e$^-$ or $\mu^+\mu^-$. In such a scenario, the observed source-to-source variation in spectral shape and intensity could be explained by variations in the galactic environment (e.g.~stellar and gas densities, and/or diffusion properties). This scenario has the additional benefit of explaining the overabundance of low latitude sources as observed by Ref.~\citep{new}, since these regions are likely to contain the highest densities of secondary targets. On the other hand, these mechanisms have difficulties in explaining the brightness of the observed continuum emission, as secondaries in small galactic substructures are unlikely to lose significant fractions of their energy before diffusing into the surrounding interstellar medium. It would thus be very unlikely in this scenario for any dark matter substructures to be bright enough to be contained within the 2FGL catalog.



\section{Discussion and Conclusions}

In this paper, we have studied the spectra, luminosities, and sky distribution of the 12 unassociated 2FGL gamma-ray sources which exhibit possible line-emission at 129 or 111 GeV, as discussed in the recent Ref.~\cite{new}. The conclusions of our study can be summarized as follows:
\begin{itemize}
\item{The majority of these 12 sources are very unlikely to be dark matter subhalos. In particular, 10 out of the 12 sources have measured spectra which are inconsistent with that predicted from dark matter annihilations. Furthermore, these sources are overwhelmingly concentrated at low galactic latitudes, in sharp contrast to that expected from a population of dark matter subhalos.}
\item{If we consider only the two 2FGL sources which exhibit spectra that are consistent with dark matter annihilation continuum emission (J0555.9-4348 and J2351.6-7558), the possibility that either or both of these sources may be subhalos cannot be ruled out. These two sources reside at moderate galactic latitudes (-40.6$^\circ$ and -27.8$^\circ$) and have overall fluxes which are consistent with the constraints on gamma-ray continuum emission.}
\end{itemize}

Looking at this problem from another perspective, the first of these two points implies that most (at least 11 out of the total 13) of the prospective line photons from unassociated 2FGL sources, as identified in Ref.~\cite{new}, do not likely originate from dark matter annihilations, and instead require some other explanation. As gamma-ray lines of uniform energy are not observed or predicted from any known class of astrophysical objects, the possibility of an astrophysical solution seems unlikely. Instead, the observation of a line-like-signal from these unassociated sources strengthens the case that the reports of gamma-rays lines over the past several months are likely to be the result of some as of yet not understood systematic effect or instrumental issue.

This hypothesis will be put to the test by future observations by other gamma-ray telescopes. Specifically, pointed observations from HESS-II~\cite{Bergstrom:2012vd} will be instrumental in determining whether the observed line signals are real or not. Additionally, the Gamma-400 instrument promises a significant improvement in both the angular and energy resolution, which will enable a detailed comparison between the tentative line feature and the continuum emission predicted in dark matter scenarios~\citep{gamma400}.



\bigskip

{\it Acknowledgements:} We would like to thank Doug Finkbeiner for valuable discussions.  DH is supported by the US Department of Energy, including grant DE-FG02-95ER40896.

\end{document}